# Proposing a fungal metabolite-Flaviolin as a potential inhibitor of 3CL$^{pro}$ of novel coronavirus SARS-CoV2 using docking and molecular dynamics

Priyashi Rao[1], Arpit Shukla[2], Paritosh Parmar[2], Dweipayan Goswami[2*]

[1]Department of Biochemistry & Forensic Science, University School of Sciences, Gujarat University, Ahmedabad 380009, Gujarat, India.

[2]Department of Microbiology & Biotechnology, University School of Sciences, Gujarat University, Ahmedabad 380009, Gujarat, India.

**Details of correspondence**

Dr. Dweipayan Goswami

Assistant Professor,

Department of Microbiology & Biotechnology,

University School of Sciences, Gujarat University,

Ahmedabad 380009, Gujarat, India.

Email: dweipayan.goswami@gujaratuniversity.ac.in

**Author Approvals**

All authors have seen and approved the manuscript, and that it hasn't been accepted or published elsewhere

**Authors have no competing interests to declare**

**Abstract**

Here after performing docking and molecular dynamics of various small molecules derived as a secondary metabolite from fungi, we propose Flaviolin to act as potent inhibitor of 3-chymotrypsin (3C) like protease (3CL$^{pro}$) of noval corona virus SARS-CoV2 responsible for pandemic condition caused by coronavirus disease 2019 (COVID-19).

The onset of 2020 witnessed the world brought to a virtual standstill with the outbreak of a novel severe acute respiratory syndrome virus (SARS-CoV-2). The coronavirus disease 2019 (COVID 2019) pandemic of the 21$^{st}$ century began as an outbreak in China in 2019 and soon, it overwhelmed the world, ultimately leading the World Health Organization to declare it as a global pandemic in March 2020. As on 28$^{th}$ March 2020, more than 500,000 confirmed cases in 202 countries were infected with the deadly SARS-CoV-2 virus with more than 23,000 deaths globally. With no apparent approved or clinically tested anti-viral drug for SARS-CoV, the race to find an efficient therapeutic has become the primary motive for the affected nations(de Wit, van Doremalen, Falzarano, & Munster, 2016; Jin, Du, Xu, Deng, Liu, Zhao, Zhang, Li, Zhang, Peng, et al., 2020; Wu et al., 2020; Zhou et al., 2020).

The genome of coronaviruses contains 25 to 32 kb, which encodes two large and overlapping polyproteins, pp1a and pp1ab, which are vital for the replication, transcription and therefore the proliferation of the virus within the host. The functional polypeptide is acted upon by virally encoded 3-chymotrypsin (3C) like protease (3CL$^{pro}$), known as main protease. Though, 3CL$^{pro}$ is a dimer, which provides the substrate binding cleft, the protease carries out its proteolytic activity by the nucleophilic attack guided by cysteine thiol in its Cys-His dyad. Owing to the vitality of 3CL$^{pro}$ in realizing the viral life cycle and the absence of its human analogue, 3CL$^{pro}$ is a prime target for the design of anti-SARS-CoV-2 drug design(Jin, Du, Xu, Deng, Liu, Zhao, Zhang, Li, Zhang, Peng, et al., 2020; Jin, Du, Xu, Deng, Liu, Zhao, Zhang, Li, Zhang, & Peng, 2020; F. Wang, Chen, Tan, Yang, & Yang, 2016).

In the present study, hundred secondary metabolites of fungi were screened virtually to assess its molecular docking with 3CL$^{pro}$ in silico. The natural chemical compounds/fungal metabolites were screened based on their *in-silico* binding energy, formation of hydrogen bonds to 3CL$^{pro}$. Flaviolin was found to be the most ideal compound to bind with 3CLpro, which was corroborated with molecular dynamics study of the flabiolin-3CL$^{pro}$ complex for 1000 ps. The stability of the complex could possibly pave a path for identifying/realizing the anti-3CL$^{pro}$ mode of action of Flaviolin. The protein 3CL$^{pro}$ was retrieved from Protein databank (PDB), ID 7LU6. The protein was co-crystallized with its cognate ligand, N3 (PubChem CID: 6323191). The co-ordinates of N3 binding site on 3CL$^{pro}$ were determined using UCSF Chimera (version 1.14). The protein structure was prepared for docking by addition of H-bonds and removal of all the co-crystallized residues (Figure 1). The energy minimization step was performed out using DockPrep tool (Krivov, Shapovalov, & Dunbrack, 2009). AM1-BCC method, known for its promptness in assigning charges to the molecules, was employed using ANTECHAMBER algorithm. while assigning the charges (J. Wang, Wang, Kollman, & Case, 2001). Energy minimization was performed using 500 steepest descent steps with 0.02 A° step size with an update interval of 10. All the steps mentioned were performed in Chimera.

The fungal secondary metabolites, ligands, were retrieved from PubChem for the *in-silico* interaction assays. Prior to the molecular docking of ligand and receptor, all ligands were optimized by addition of hydrogen and energy minimization using Gasteiger algorithm (Gasteiger & Jochum, 1979) in structure editing wizard of Chimera, which works on the chemoinformatic principle of electronegativity equilibration and the files were saved in mol2 format.

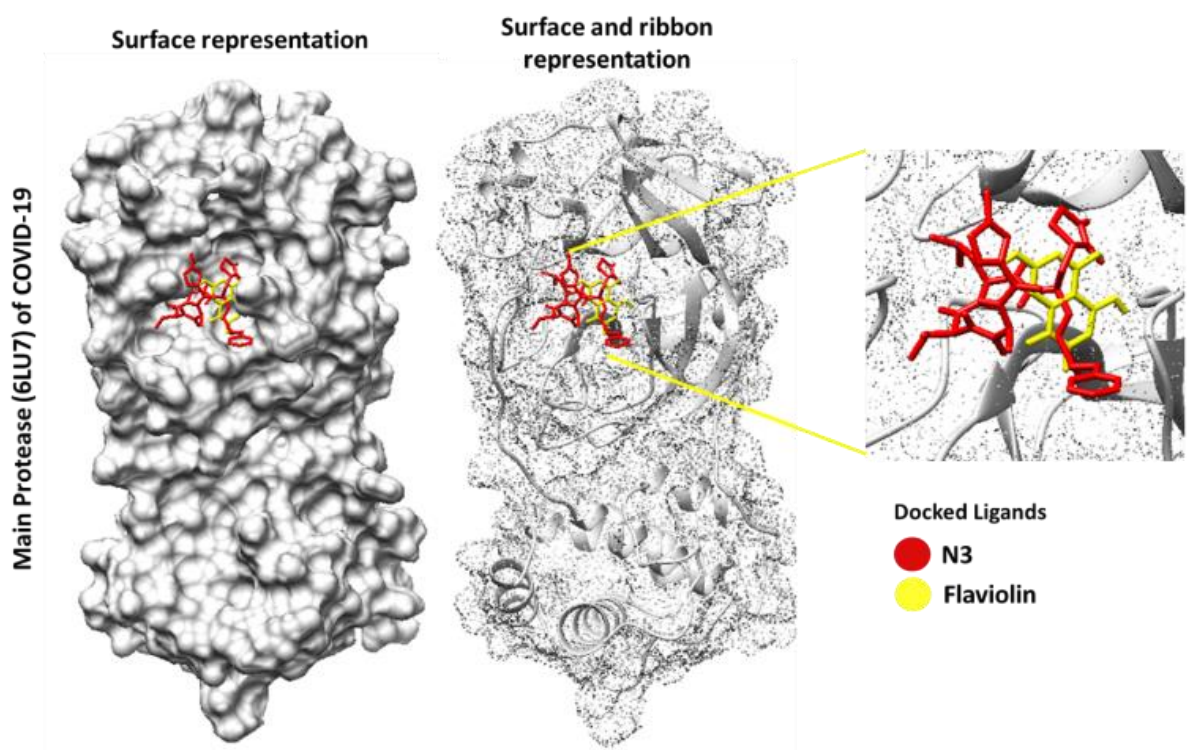

**Figure 1** 3D image representing binding of N3 and Flaviolin at same cleft of 3CL$^{pro}$ (PDB ID: 6LU7) of COVID-19.

Receptor-ligand docking analysis was performed using AutoDock Vina (Trott & Olson, 2010) and the program was executed as an add-on in Chimera. The ligand binding site in 3CL$^{pro}$ was selected based on the crystallized cognate ligand (N3), attached in the original pdb file and the co-ordinates were recorded for docking of fungal ligands. Further, the coordinates of hydrophobic cavity of 3CL$^{pro}$ active site was used in the docking of N3 (as control) and the fungal metabolites.

In the AutoDock Vina algorithm, the following parameters were set as: (i) number of binding modes- 10; (ii) exhaustiveness of search- 8 and (iii) maximum energy difference- 3 kcal/mol. Out of all the possible poses suggested by AutoDock Vina, the pose showing maximum hydrogen bonds and minimum binding free energy change (kcal/mol), as represented in the ViewDock window, were chosen. They were further analysed in Biovia Discovery Studio (DS) visualizer for the formation of hydrogen bonds, in addition to other supporting hydrophobic interactions, by the functional groups of ligands with the amino acids of 3CL$^{pro}$. The metabolite (ligand) making maximum number of H-bonds, showing capability to form covalent interaction with 3CL$^{pro}$ and showed highest binding affinity was chosen for further study using molecular dynamics simulation.

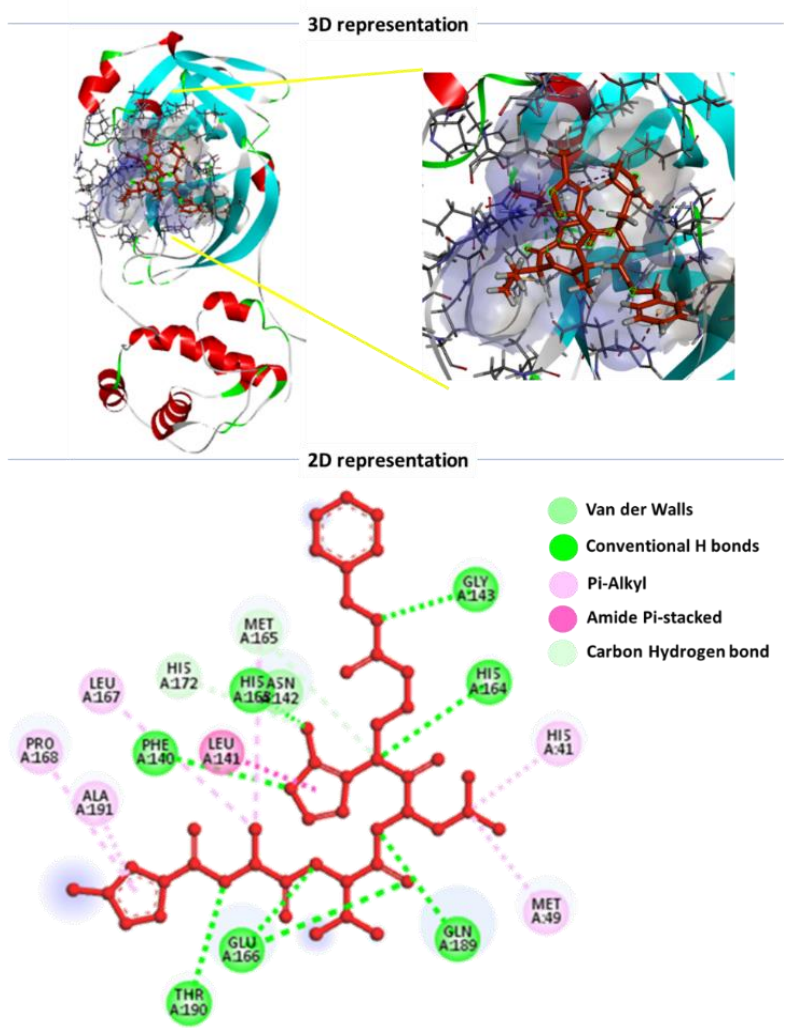

**Figure 2** Interaction of N3 in the binding cleft of 3CL$^{pro}$ (PDB ID: 6LU7) of COVID-19 shown in 3D representation and 2D representation describing ligands interactions by formation of various H-bonds and hydrophobic interactions with protein.

The structure analysis of SARS-CoV2 3CL$^{pro}$ shows that, the substrate binding site is located in the Cys-His catalytic dyad. Domain I extend from residues 8 to 101 and Domain II ranges from residues 102 to 184 comprise of antiparallel β-barrel structure. Domain residues is formed from residue 201 to 303 and possess five α-helices arranged into a largely antiparallel globular cluster, which is connected to domain II means of a long loop region extending from residue 185 to 200. For COVID-19, the substrate binding site is located in the cleft between Domain I and II. Binding of N3 with 3CL$^{pro}$ shows that it binds in the substrate binding pocket in (Figure 1). N3 is relatively a large molecule where its backbone forms an antiparallel sheet with residues 164 to168 (His164, Glu166, Met165, Leu167, Pro168) of 3CL$^{pro}$ while also interacting with residues 189 to191 (Gln189, Thr190, Ala191) of the loop that links Domain II and III.

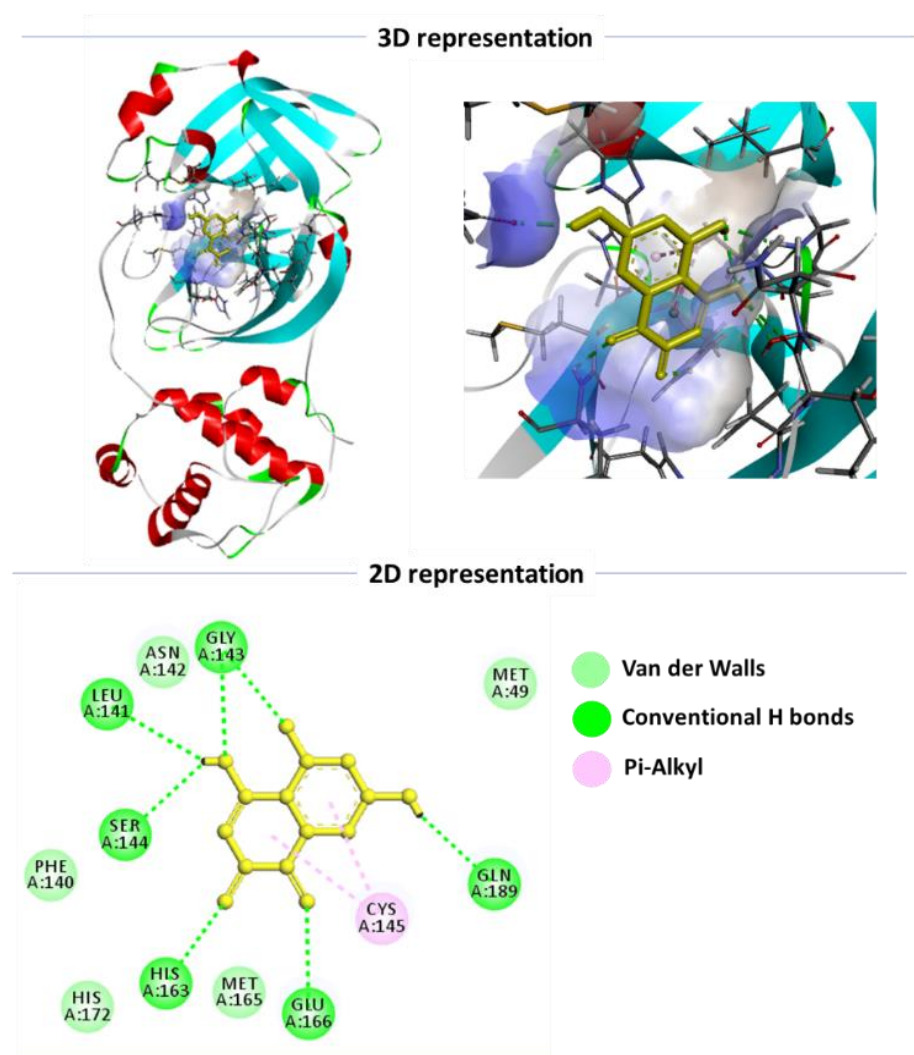

**Figure 3** Interaction of Flaviolin in the binding cleft of 3CL$^{pro}$ (PDB ID: 6LU7) of COVID-19 shown in 3D representation and 2D representation describing ligands interactions by formation of various H-bonds and hydrophobic interactions with protein.

The group of scientists who co-crystalized this protein with N3 also suggest strong covalent bond formed by Cys145 of 3CL$^{pro}$ with N3 of the 1.8 Å length, in turn, making the interaction with of this inhibitor bind irreversibly with the protein. Briefly, the inhibitor N3 makes six hydrogen bonds (with Gly143, Phe140, His163, His164, Glu166 and Thr190), one amide pi-stacked interaction (with Leu141), five pi-alkyl interactions (with His41, Met49, Leu167, Pro168 and Ala191) two carbon hydrogen bonds (Met165 and His172) and one van der Walls interactions (with Asn142) with 3CL$^{pro}$. For the docking of fungal metabolites with 3CLpro, the coordinates of N3 binding were used to assess the binding affinity of various fungal metabolites with 3CL$^{pro}$. On performing the docking of all the metabolites one by one with 3CL$^{pro}$, it was observed that, only five molecules were able to make three hydrogen bonds or more with 3CL$^{pro}$ at the binding cleft of N3. These were Asperic acid, Aspernigrin A, Aurosperone B, Carbomymethyl-3-hexylmaleic acid and the best one, flaviolin. Flaviolin showed to make eight hydrogen bonds in total (with Leu141, Gly143, Ser144, Cys145, His163, Glu166 and Gln189) and made one carbon hydrogen bond (with Asn142) and the docking energy as predicted by AutoDock Vina was -7.3 kcal/mol for the best pose (Figure 2). As shown previously, N3 showed interaction with residues 164 to168 (His164, Glu166, Met165, Leu167, Pro168) of 3CL$^{pro}$ along with interacting with residues 189 to191 (Gln189, Thr190, Ala191) of the loop that links Domain II and III, flaviolin also showed similar interaction with 3CL$^{pro}$. All these properties make flaviolin as the best choice as the most suitable inhibitor for 3CL$^{pro}$. When the docked best pose of flaviolin was made to superimpose with that of N3 on 3CLpro, it was observed that it was able to bind exactly at the same N3 binding cleft (Figure 1). Moreover, the docking analysis suggested that flaviolin interacted with almost all similar amino acids akin to N3-3CL$^{pro}$ binding.

Further, to validate the results of docking, the comparative molecular dynamics (MD) simulation was performed in two pairs of experiments studying (i) N3-3CL$^{pro}$ and (ii) Flaviolin-3CL$^{pro}$ individually and their results were compared. Steepest descent energy minimization was performed for both the systems, and the systems were equilibrated under NVT (constant number of particles, volume and temperature) conditions for 1000 ps at 300 K, and the MD run was initiated. Once the NVT run was completed the system was proceeded with NPT (constant number of particles, pressure, and temperature) simulation and MD run was performed for 1000 ps. Trajectories were analysed for root-mean-square deviation (RMSD), root-mean-square fluctuation (RMSF), radius of gyration (Rg) and the number of H-bonds formed between the ligand and proteins using 'gmx rms',' gmx rmsf', 'gmx gyrate' and 'gmx hbond' of GROMACS utilities (Bekker et al., 1993; Pronk et al., 2013). Ligand–protein stability was determined by the dynamics of hydrogen bonds between ligand and protein with respect to time. XMgrace tool was used to prepare the graphs (Turner, Land-Margin Research, & Technology, 2005).

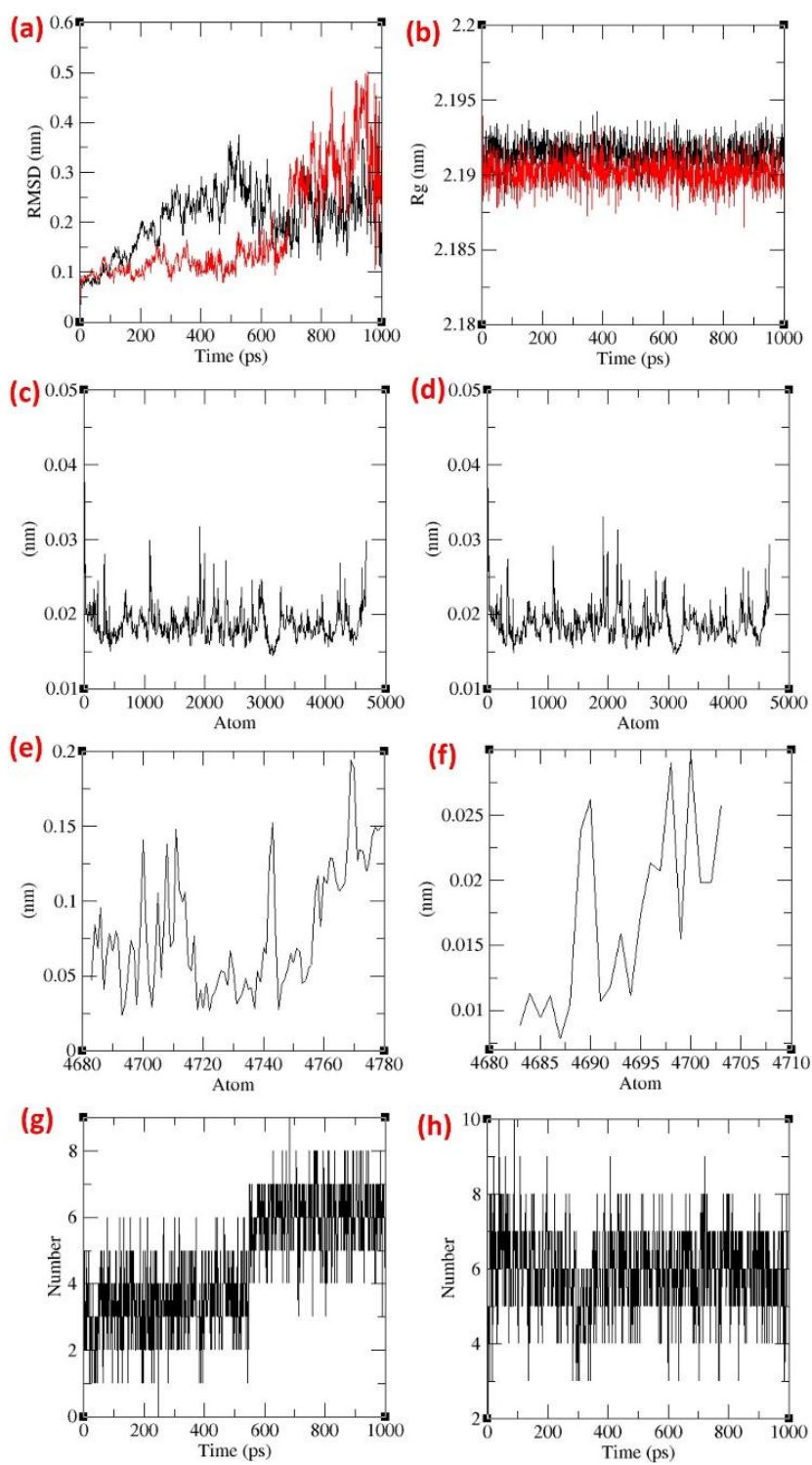

**Figure 4** Representation RMSDs of (a) 3CL$^{pro}$ backbone RMSD during interaction with N3 (black) and Flaviolin (red), (b) Radius of gyration of Cα atoms of 3CL$^{pro}$ of COVID-19 in the presence of N3 (black) and Flaviolin (red), (c) RMSF values of 3CL$^{pro}$ backbone during its interaction with N3, (d) RMSF values of 3CL$^{pro}$ backbone during its interaction with Flaviolin (e) ligand N3's RMSF values during simulation, and (f) ligand Flaviolin's RMSF values during simulation; Hydrogen Bonds formed by (g) N3 and (h) Flaviolin with 3CL$^{pro}$ of COVID-19 during simulation.

The RMSD values for the backbone of proteins in presence of both the ligands, 3CL$^{pro}$ was found to be more stable in presence of flaviolin as compared to N3. Though, on stabilization of the system after 500 ps of MD run, the RMSD values were almost equal (Figure 4a). Radius of gyration (Rg) was determined to understand the level of compaction in the structure of 3CL$^{pro}$ in the presence of N3 and Flaviolin. The Rg is defined as the mass weighted root mean square distance of a collection of atoms from their common center of mass. Hence this analysis gives us the overall dimensions of the protein. The calculated Rg values over the simulation time scale for (i) N3-3CL$^{pro}$ and (ii) Flaviolin-3CL$^{pro}$ are shown in Figure 4b. The Rg value of N3-3CL$^{pro}$ and Flaviolin-3CL$^{pro}$ varies between 2.192 nm and 2.190 nm. From the Rg plot, it shows that Flaviolin-3CL$^{pro}$ has marginally lower Rg values, however the difference cannot be considered significant. This suggest the protein behave identical in presence of N3 and Flaviolin.

The RMSF with respect to the average MD simulation conformation reflects as a mean portraying flexibility differences among residues. The RMSF of the backbone atoms of each residue in the (i) N3-3CL$^{pro}$ and (ii) Flaviolin-3CL$^{pro}$ individually is calculated to reveal the flexibility of the backbone structure in presence of both ligands. This shows how the protein behaves in presence of different ligands. The high RMSF value indicates more flexibility whereas the low RMSF value indicates limited movements during simulation in relation to its average position. The RMSF of the protein backbone 3CL$^{pro}$ in presence of both the ligands N3 and Flaviolin were almost identical suggesting similar affinity of binding with the protein. If the binding is poor than the protein is freer to hive higher RMSF values, which is not the case here (Figure 5c, d). While the RMSF of the ligands N3 and Flaviolin is shown in Figure 5e and f respectively. As both the ligands are differing in size and shapes, comparing their RMSF doesn't hold any importance.

The intermolecular hydrogen bonding between the protein and the ligand plays an essential role in stabilizing the protein–ligand complexes. The stability of the hydrogen bond network formed between (i) N3-3CL$^{pro}$ and (ii) Flaviolin-3CL$^{pro}$ were calculated throughout the simulation at 300 K for the ligated system and the results are depicted in Figure 5e and f. The N3-3CL$^{pro}$ complex exhibited four hydrogen H-bonds upto 500ps and thereafter the number of hydrogen bond increased to the average of six suggesting the ligand getting more stable during the course of MD run. While for Flaviolin-3CLpro, there were consistently six hydrogen bonds being formed throughout the simulation. Overall scenario suggests both ligands fits in to the binding cleft making appropriate hydrogen bonds.

From the MD analysis it can be said that the protein behaves identical in presence of both the ligands, N3 and Flaviolin both forms similar hydrogen bonds with 3CL$^{pro}$ and hence the efficacy of Flaviolin to interact with 3CL$^{pro}$ can be depicted to be at par with N3.

**Table 1** ADMET prediction of N3 an Flavolin.

| Model Name | N3 Predicted Value | Flaviolin Predicted Value | UNIT |
|---|---|---|---|
| Water solubility | -4.181 | -2.042 | (log mol/L) |
| Caco2 permeability | 0.501 | 0.215 | (log Papp in $10^{-6}$ cm/s) |
| Intestinal absorption | 62.398 | 62.173 | (% Absorbed) |
| Skin Permeability | -2.736 | -2.771 | (log Kp) |
| P-glycoprotein substrate | Yes | No | (Yes/No) |
| P-glycoprotein I inhibitor | Yes | No | (Yes/No) |
| P-glycoprotein II inhibitor | No | No | (Yes/No) |
| VDss (human) | -0.764 | 0.313 | (log L/kg) |
| Fraction unbound (human) | 0.052 | 0.685 | Numeric (Fu) |
| BBB permeability | -1.725 | -0.776 | (log BB) |
| CNS permeability | -4.013 | -3.294 | (log PS) |
| CYP2D6 substrate | No | No | (Yes/No) |
| CYP3A4 substrate | Yes | No | (Yes/No) |
| CYP1A2 inhibitor | No | No | (Yes/No) |
| CYP2C19 inhibitor | No | No | (Yes/No) |
| CYP2C9 inhibitor | No | No | (Yes/No) |
| CYP2D6 inhibitor | No | No | (Yes/No) |
| CYP3A4 inhibitor | Yes | No | (Yes/No) |
| Total Clearance | 0.713 | 0.425 | (log ml/min/kg) |
| Renal OCT2 substrate | No | No | (Yes/No) |
| AMES toxicity | No | No | (Yes/No) |
| Max. tolerated dose (human) | 0.015 | 0.29 | (log mg/kg/day) |
| hERG I inhibitor | No | No | (Yes/No) |
| hERG II inhibitor | Yes | No | (Yes/No) |
| Oral Rat Acute Toxicity (LD50) | 4.138 | 1.544 | (mol/kg) |
| Oral Rat Chronic Toxicity (LOAEL) | 3.606 | 2.688 | (log mg/kg_bw/day) |
| Hepatotoxicity | Yes | No | (Yes/No) |
| Skin Sensitisation | No | No | (Yes/No) |
| *T. Pyriformis* toxicity | 0.285 | 0.288 | (log ug/L) |
| Minnow toxicity | 4.885 | 2.952 | (log mM) |

The pkCSM - pharmacokinetics server (Pires, Blundell, & Ascher, 2015) was used to predict the ADMET (absorption–distribution–metabolism–excretion–toxicity) properties of N3 and flaviolin. It predicted both physiochemical and pharmacological properties. SMILES (Simplified Molecule Input Line Entry Specification) of the compounds were retrieved from PubChem and uploaded to pkCSM - pharmacokinetics server (Table 1). Of all the tests, Flaviolin showed better distribution based on the values of volume of Distribution (VDss). Flaviolin also showed fraction unbounded showing its ability to distribute in tissues of body more effectively. Flaviolin does not even interfere the Cytochrome (CYP) and has better renal clearance. Lastly the toxicity by Flaviolin on liver cells is also not predicted and hence it has better drug like properties when compared to N3. With efforts been made to control the pandemics of SAR-CoV2 in the form of novel coronavirus, to find the lead compound against the target protein that can suppress its.

Reflecting back in the history, fungal metabolites have been a boon for the mankind, starting from antibiotics to flavoring agents to food preservatives. Fungi are known to produce large amounts of secondary metabolites and by chance that might interact with 3CL$^{pro}$, with this rationale the fungal metabolites were looked for pursuing this research. Previously, the potentials of fungal metabolites as anti-viral agents are explored and the success was promising. The anti-viral compounds from fungal origin is vividly described by Linnakoski et al. (2018) where the promising anti-viral compounds portrayed were belong to the chemical class of Indole alkaloids, Non-ribosomal peptides, Polyketides, Terpenoids. With efforts been made to control the pandemic of SAR-CoV2 in the form of novel Coronavirus, to find a lead compound against the target protein that can suppress its. We here are proposing Flaviolin to interact with one of the important target proteins of SAR-CoV2, 3CL$^{pro}$ and block its function as deduced using docking and molecular dynamics.

**Acknowledgements**

Authors are thankful to Gujarat University for providing necessary facilities to perform experiments.

**Compliance with ethical standards**

This article does not contain any studies with human participants or animals performed by any of the authors.

**Conflict of interest**

The authors declare that they have no conflict of interest.